\documentclass[%
 reprint,
superscriptaddress,
nofootinbib,
 amsmath,amssymb,
 aps,
 pra,
]{revtex4-1}
\usepackage{verbatim}
\usepackage{mathtools}
\usepackage{amsthm}
\usepackage{physics}
\usepackage{graphicx}
\usepackage{dcolumn}
\usepackage{bm}
\usepackage{xfrac}
\graphicspath{ {./images/} }

\begin{document}


\title{Correlation energy and quantum correlations in a solvable model}

\author{Javier Faba}

\affiliation{%
Center for Computational Simulation, Universidad Polit\'ecnica de Madrid, Campus Montegancedo, 28660 Boadilla del Monte, Madrid, Spain
}%

\affiliation{%
Departamento de F\'isica Te\'orica and CIAFF, Universidad Aut\'onoma de Madrid, E-28049 Madrid, Spain 
}%

\author{Vicente Mart\'\i n}

\affiliation{%
Center for Computational Simulation, Universidad Polit\'ecnica de Madrid, Campus Montegancedo, 28660 Boadilla del Monte, Madrid, Spain
}%

\author{Luis Robledo}
\affiliation{%
Departamento de F\'isica Te\'orica and CIAFF, Universidad Aut\'onoma de Madrid, E-28049 Madrid, Spain 
}%

\affiliation{%
Center for Computational Simulation, Universidad Polit\'ecnica de Madrid, Campus Montegancedo, 28660 Boadilla del Monte, Madrid, Spain
}%

\date{\today}

\begin{abstract}
Typically in many-body systems the correlation energy, which is defined as the difference between the exact ground state energy and the mean-field solution, has been a measure of the system's total correlations. However, under the quantum information context, it is possible to define some quantities in terms of the system's constituents that measure the classical and quantum correlations, such as the entanglement entropy, mutual information, quantum discord, one-body entropy, etc. In this work, we apply concepts of quantum information in fermionic systems in order to study traditional correlation measures (the relative correlation energy) from a novel approach. Concretely, we analyze the two and three level Lipkin models, which are exactly solvable (but non trivial) models very used in the context of the many-body problem.
\end{abstract}

\pacs{Valid PACS appear here}
\maketitle


\section{\label{sec:introduction}Introduction}

The atomic nucleus is a mesoscopic system made of  protons and 
neutrons with strong interactions among its constituents. Due to the complexity of the nuclear interaction and the 
large number of particles involved, the dynamic governing the nucleus is
very rich, giving rise to a huge amount of different situations involving single particle and/or collective excitations \cite{ring_schuck,Bender2003}. 
As a consequence of the underlying mean field which implies the existence of well defined orbits, low energy nuclear properties 
can dramatically change
by changing a few units of the nucleus' proton and neutron numbers as a consequence of the filling of different orbits. 
At low excitation energies, the so-called 
collective excitations show more regular patterns than the single particle ones. The reason it that they are associated 
to more macroscopic-like degrees of freedom as the 
shape of the nucleus and are intimately connected with the mechanism of 
spontaneous symmetry breaking and symmetry restoration \cite{Robledo2019,Sheikh2019}. In finite systems
this mechanism can be viewed as an artifact of the underlying mean field description to 
capture correlations in a simple way. Nevertheless the breaking for symmetries at the mean field level 
is intimately connected to properties of the exact wave functions of the system. 
The subsequent symmetry restoration of the symmetry-broken mean-field wave functions gives rise to collective bands (being 
rotational bands the most prominent example) that represent a prominent part of the nuclear spectrum with very specific and
universal properties like the $I(I+1)$ energy rule of rotational bands \cite{Bender2003,Robledo2019,Sheikh2019}. 
To improve upon the mean field plus symmetry 
restoration paradigm, one usually add an additional layer where 
fluctuations on the collective degrees of freedom are explicitly 
treated. This is usually done in the framework of the Generator 
Coordinate Method \cite{Bender2003,Robledo2019}. A question that arises very often is how to quantify
the balance between the correlations associated to symmetry restoration and 
quantum fluctuations. The answer to this question might help to devise
new approaches to solve the nuclear many body problem. On the other hand,
the connection between the exact solution of the problem and the approximate
mean field plus symmetry restoration plus fluctuations approach is not
straightforward and there has been quite a lot of work to extract from the 
exact shell model solution \cite{Caurier2005} the underlying symmetry breaking mean field. 
Therefore, it is also interesting to find a quantity to be computed with 
the exact solution of the problem that is able to pin-point the quantum
phase transitions observed in the mean field description of the nucleus. This is an approach also pursued in other fields like 
quantum chemistry \cite{Legeza2006,szalay}, superconductors in condensed matter \cite{Zeng2014}, atomic physics \cite{tichy2011} and even nuclear physics \cite{yoshiko,Kruppa_2021}.
With these two goals in mind we analyze in this paper some quantum information related quantities as the overall entropy
expressed in the basis of the natural states, the quantum discord and the correlation energy. We will carry out
our study in the realm of a simple, albeit rich, exactly solvable
nuclear physics problem: the Lipkin model with two \cite{lipkin} and three \cite{Li1970,Holzwarth1974} active orbits. Both 
models show quantum phase transitions as a function of the interaction parameter strength that mimic the
spontaneous symmetry breaking mechanism discussed above.

\section{\label{sec:theory}Theoretical background}

In this section we will introduce briefly some concepts that we will use in the next sections. When dealing with correlations in a many-body system, one has to clarify two fundamental issues: what are we defining as subsystem, and how to quantify the correlations among them? If our Hilbert space is defined as a tensor product of Hilbert spaces, then the notion of subsystem arises naturally. For example, the Hilbert space of a system formed by $N$ qubits is simply the tensor product of each qubit's Hilbert space. However, if we are dealing with indistinguishable particles (fermions in our case) in the context of second quantization, we cannot define the Hilbert space as a tensor product of each particle's Hilbert space because of the (anti)symmetry of the wavefunction. A lot of effort has been made to disentangle the correlations associated to the symmetrization principle or super-selection rules from the ones coming from the dynamic of the system \cite{Legeza2003,Banuls2007,Ding2021} and quantities like the fermionic partial trace between modes \cite{fermionic_partial_trace} or the von Neumann entropy of the one-body density matrix \cite{yoshiko} have been defined. Those quantities have been thoroughly used in the literature \cite{Legeza2003,Legeza2006,szalay,gigena_overall_entropy,Kruppa_2021,CRobin}. In this work, we will discuss quantities that make use of both concepts. 

There are many possibilities in order to characterize and quantify correlations in a quantum system. Typically, if our Hilbert space can be written as $\mathcal{H} = \mathcal{H}_A\otimes \mathcal{H}_B$\footnote{As discussed above, when dealing with indistinguishable particles in the second quantization formalism we don't have a tensor product structure. However, if we define the subsystems as the single particle states (also called orbitals or modes in this work), we can treat the system as a tensor product if we take into account some subtleties which arise from the fermionic anticommutation rules \cite{fermionic_partial_trace}.} we can measure the entanglement between the $A$ and $B$ subsystems for a given pure state $\ket{\psi}\in\mathcal{H}$ through the von Neumann entropy of the reduced states, namely:
\begin{equation}
\label{eq:vn_entropy}
    S(\rho^{(A)}) = -\Tr(\rho^{(A)}\ln\rho^{(A)})
\end{equation}
where $\rho^{(A)} = \Tr_B(\ket{\psi}\bra{\psi})$ \cite{nielsen_chuang}. However, if we are dealing with mixed states, this method is no longer valid as an entanglement measure. Furthermore, entanglement is not only the only type of correlation present in a quantum system: it can also have classical correlations, and quantum correlations beyond entanglement.

The quantum discord \cite{quantum_discord} is a measurement-based quantity of the total quantum correlations (including entanglement and beyond) between two subsystems. It is defined as
\begin{equation*}
\delta(A,B) = I(A,B) - J(A,B).
\end{equation*}
where $I(A,B) = S(\rho^{(A)})+S(\rho^{(B)})-S(\rho^{(A,B)})$ is the mutual information, and $J(A,B)$ is defined as
\begin{equation}
\label{eq:J}
J(A,B) = \max_{\{\Pi_{k}^{(B)}\}} S(\rho^{(A)})-S(\rho^{(A,B)}|\{\Pi_{k}^{(B)}\}).
\end{equation}
While $I(A,B)$ is a measure of all kind of correlations, $J(A,B)$ quantifies only the classical part. The measurement-based conditional entropy in Eq. (\ref{eq:J}) is defined as
\begin{equation*}
S(\rho^{(A,B)}|\{\Pi_{k}^{(B)}\}) = \sum_{k} p_k S(\rho_k^{(A,B)})
\end{equation*}
where $\rho_k^{(A,B)} = \frac{1}{p_k}\Pi_{k}^{(B)}\rho^{(A,B)}\Pi_{k}^{(B)}$ is the measured-projected total state and $p_k = \tr (\Pi_{k}^{(B)}\rho^{(A,B)}\Pi_{k}^{(B)})$ is the associated probability. The measurement and the associated projector $\Pi_{k}^{(B)}$ are defined only in the sector $B$ of the bi-partition. For pure states, the quantum discord reduces to the entanglement between subsystems with $J(A,B) = \delta(A,B)$ \cite{luo}. However, for mixed states this is not true in general. This quantity is interesting since, as we will see in the next sections, it can be a useful measure in order to study quantum phase transitions in many-body systems \cite{discord_phase_transitions,discord_phase_transitions_2,discord_phase_transitions_3}.

However Eq. (\ref{eq:J}) requires a variational procedure involving all possible $B$-subsystem projectors, so that computing quantum discord is in general analytically and computationally intractable \cite{QD_is_NP}. Fortunately, if we are dealing with fermionic systems, no optimization process is needed in order to compute quantum discord between two arbitrary orbitals \cite{two_orbital_quantum_discord}.

Other useful measure of system's correlations is the overall entropy, defined as
\begin{equation*}
    S_{ov} = \sum_i S(\rho^{(i)})
\end{equation*}
where $\rho^{(i)}$ is the reduced density matrix for the $i$-th orbital. Its value is a measure of the total system's correlations, if the total state is pure \cite{szalay}. It is closely related to the one-body entropy, defined as the von Neumann entropy of the one-body density matrix \cite{yoshiko}, whose elements are $\gamma_{ij} = \langle c^\dagger_jc_i\rangle$. Because of the parity super-selection rule \cite{PSSR} we have
\begin{equation*}
    \rho^{(i)} = 
    \begin{pmatrix}
        1-\langle c^\dagger_ic_i\rangle & 0 \\
        0 & \langle c^\dagger_ic_i\rangle \\
    \end{pmatrix}
\end{equation*}
where the operator $c^\dagger_i$ ($c_i$) creates (annihilates) a particle in the $i$-th orbital and the usual fermionic anticommutation rules $\{c^\dagger_i,c_j\} = \delta_{ij}$, $\{c_i,c_j\} = 0$ are fulfilled. If the overall entropy is evaluated in the natural orbital basis $\{a^\dagger_i\}$ (which is defined as the one that diagonalizes the one-body density matrix, and it has been shown that is the basis that minimizes the overall entropy \cite{gigena_overall_entropy}), then
\begin{equation}
\label{eq:overall_entropy_and_onebody_entropy}
\begin{split}
    S_{ov}^{nat} &= \sum_i f(\langle a^\dagger_ia_i\rangle) \\
    S(\gamma)&= \sum_i g(\langle a^\dagger_ia_i\rangle)
\end{split}
\end{equation}
where the functions $f$ and $g$ are defined as $f(x) =-(1-x)\log(1-x)-x\log(x) $ and $g(x) =-x\log(x) $. Since both $f$ and $g$ satisfy $f(0)=g(0)=f(1)=g(1)=0$ and are real valued smooth and strictly concave functions, the information and behaviour of $S_{ov}^{nat}$ and $S(\gamma)$ are essentially the same.

As we will discuss in the following sections it will be useful to compare this quantity, which quantifies the total system correlation (under a quantum information perspective), with the relative correlation energy \cite{corr_energy_1,corr_energy_2}, defined as
\begin{equation*}
    \epsilon_{corr} = \frac{E_{exact}-E_{HF}}{E_{exact}}
\end{equation*}
where $E_{exact}$ is the exact ground state energy and $E_{HF}$ is the  ground state energy obtained at the mean field (HF) level. Traditionally, $\epsilon_{corr}$ has been used to quantify the amount of correlations in a system, since it compares the exact ground state energy which contains all the correlations in the system with the mean field one which is taken here as an uncorrelated reference. Moreover, the correlation energy is closely related with the overlap between the exact ground state and the Hartree-Fock one \cite{Benavides-Riveros}.

\section{\label{sec:2_level_lipkin}Two-level Lipkin model}

In this section we will discuss the quantities previously defined, under the context of the two-level Lipkin model.

The so called Lipkin model \cite{lipkin} (proposed by Lipkin, Meshkov 
and Glick in 1964) consists of a $N$-fermion two level system separated 
by an energy gap $\epsilon$, each level having a $N$-fold degeneracy (we 
assume that all fermions are of the same type and have no spin, for 
simplicity). We label the upper/lower level with the quantum number 
$\sigma = +$ or $\sigma = -$ respectively, and the degeneracy with the 
quantum number $p = 1,2, \ldots, N$. The quantum number $\sigma$ can also be interpreted as a
parity quantum number (see below). The Hamiltonian is given in terms of fermionic creation and 
annihilation operators by
\begin{equation}
\label{eq:2_Lipkin_Hamiltonian}
H = \epsilon K_0 - \frac{1}{2}V(K_+K_++K_-K_-)
\end{equation}
with
\begin{equation*}
\begin{split}
K_0 &= \frac{1}{2}\sum_{p=1}^N(c^\dagger_{+,p}c_{+,p}-c^\dagger_{-,p}c_{-,p}) \\
K_+ &= \sum_{p=1}^Nc^\dagger_{+,p}c_{-,p} \qquad K_- = (K_+)^\dagger 
\end{split}
\end{equation*}
As the interaction is of the
monopole-monopole type, the quantum number $p$ is conserved in the model.

The advantage of this model is that it is exactly solvable, since the operators introduced in Eq. (\ref{eq:2_Lipkin_Hamiltonian}) are the generators of the algebra of $SU(2)$\footnote{See Refs. \cite{robledo,entanglement_lipkin_model} and references therein for a detailed discussion of the exact solution.}. 

\begin{figure}
\includegraphics[width=0.5\textwidth]{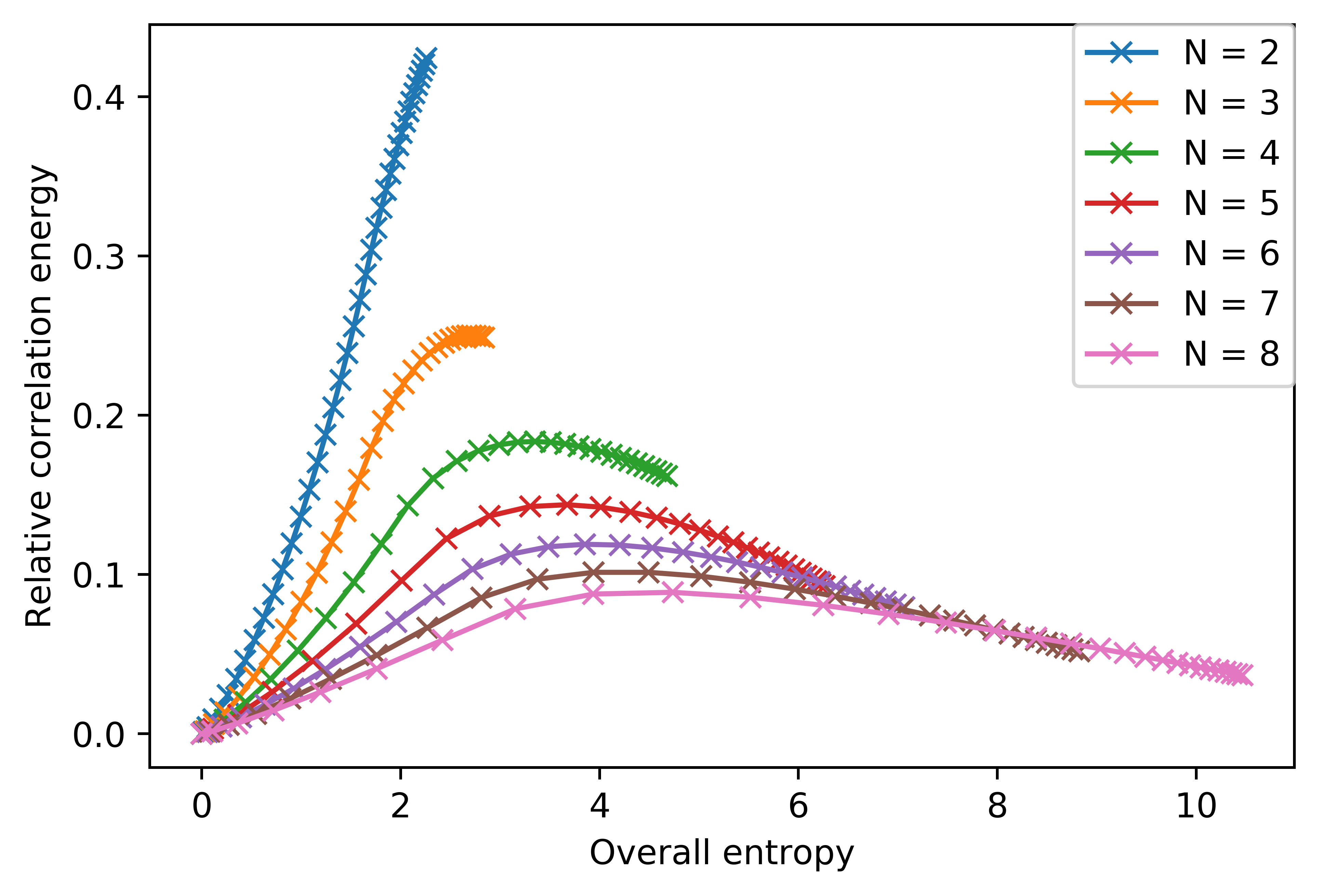}
\caption{Relative correlation energy as a function of the overall entropy, for the exact ground state of the two level Lipkin model and different particle number $N$.}
\label{fig:2_lipkin_corr_energy_vs_overall_entropy}
\end{figure}

The mean field (HF) solution can be easily obtained \cite{lipkin} because the HF energy depends on a single variational parameter. Defining the 
dimensionless interaction strength $\chi = \frac{(N-1)V}{\epsilon}$ it is observed that, for certain values of $\chi$ the HF solution breaks the parity symmetry of the Hamiltonian in Eq. (\ref{eq:2_Lipkin_Hamiltonian}) (to be associated with the $\sigma$ quantum number). With the above definitions,  the parity operator is defined as
\begin{equation*}
    P_z = \exp(i\pi\sum_{p=1}^{N}c^\dagger_{-,p}c_{-,p}).
\end{equation*}
The HF states will have a well defined parity if they are eigenstates of the parity operator $P_z$.
When $\chi\leq 1$ the HF solution preserves the parity symmetry (spherical phase) whereas the symmetry is broken (deformed phase) when $\chi>1$  \cite{robledo}. 
The correlation energy of the ground state can be easily computed by comparing the exact solution with the Hartree-Fock (HF) solution. 
This quantity as well as the the overall entropy in the natural orbital basis depend on the strength parameter $\chi$ and they are strongly correlated as can be seen in Fig. \ref{fig:2_lipkin_corr_energy_vs_overall_entropy}.

\begin{figure}
\includegraphics[width=0.5\textwidth]{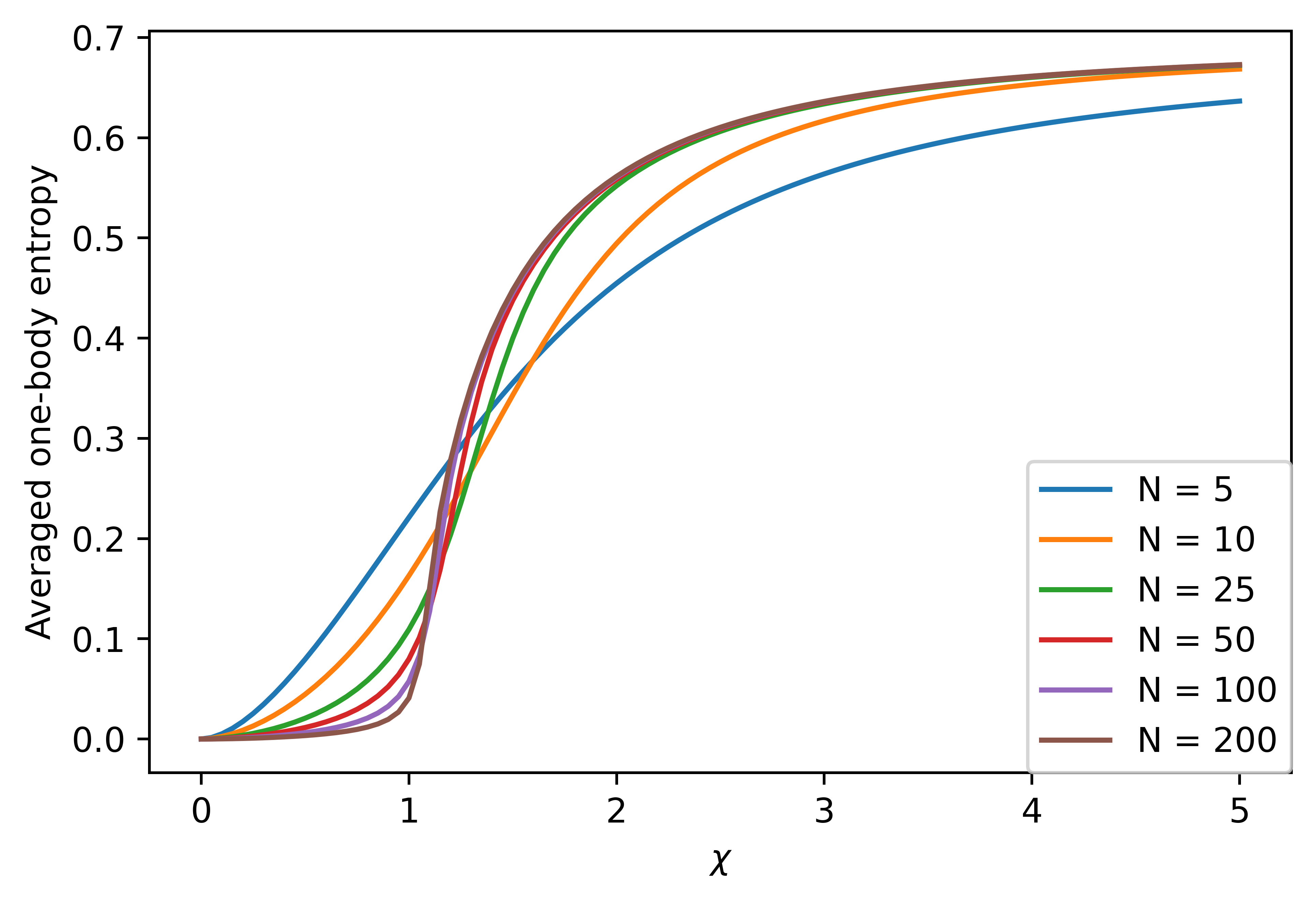}
\caption{Averaged one body entropy of the exact ground state as a function of the parameter $\kappa$. As the number of particles increases, the transition between the spherical phase (low correlation regime) and the deformed one (high correlated regime) is sharper.}
\label{fig:2_lipkin_onebody_vs_kappa}
\end{figure}

We can distinguish three regions in this plot. For low enough values of the overall entropy, the relative correlation energy grows quasi-linearly. Then, after a sudden discontinuity of the second derivative (see Fig. \ref{fig:second_derivative_2_lipkin_corr_energy_vs_overall_entropy} below) the relative correlation energy reaches a maximum and then bends down to gently decrease until the overall entropy saturates. This change of tendency is due to the phase transition observed in the HF solution at $\chi=1$. Fig. \ref{fig:2_lipkin_corr_energy_vs_overall_entropy} can also be interpreted in terms of the values of $\chi$. In the spherical phase ($\chi\leq1$) the mean field solution catches as many correlations as possible while preserving the non-interacting picture and preserving the system's symmetry. As the correlation/interaction grows (quantified by the overall entropy/the parameter $\chi$) the relative correlation energy grows too, showing that the mean field approach is less accurate since the difference between $E_{HF}$ and $E_{exact}$ is getting bigger. When $\chi \geq 1$, the system's correlations are too strong and the mean field solution breaks the parity symmetry in order to catch as much as possible of them (see Fig. \ref{fig:2_lipkin_onebody_vs_kappa}). In this way, the relative correlation energy shows a decreasing behaviour until the saturation of the overall entropy. This change in the behaviour of the system can also be seen in Fig. \ref{fig:second_derivative_2_lipkin_corr_energy_vs_overall_entropy}, where the second derivative of the relative correlation energy is plotted as a function of the overall entropy. A sudden jump is observed in this quantity when $\chi=1$ signaling the quantum phase transition.

\begin{figure}
\includegraphics[width=0.5\textwidth]{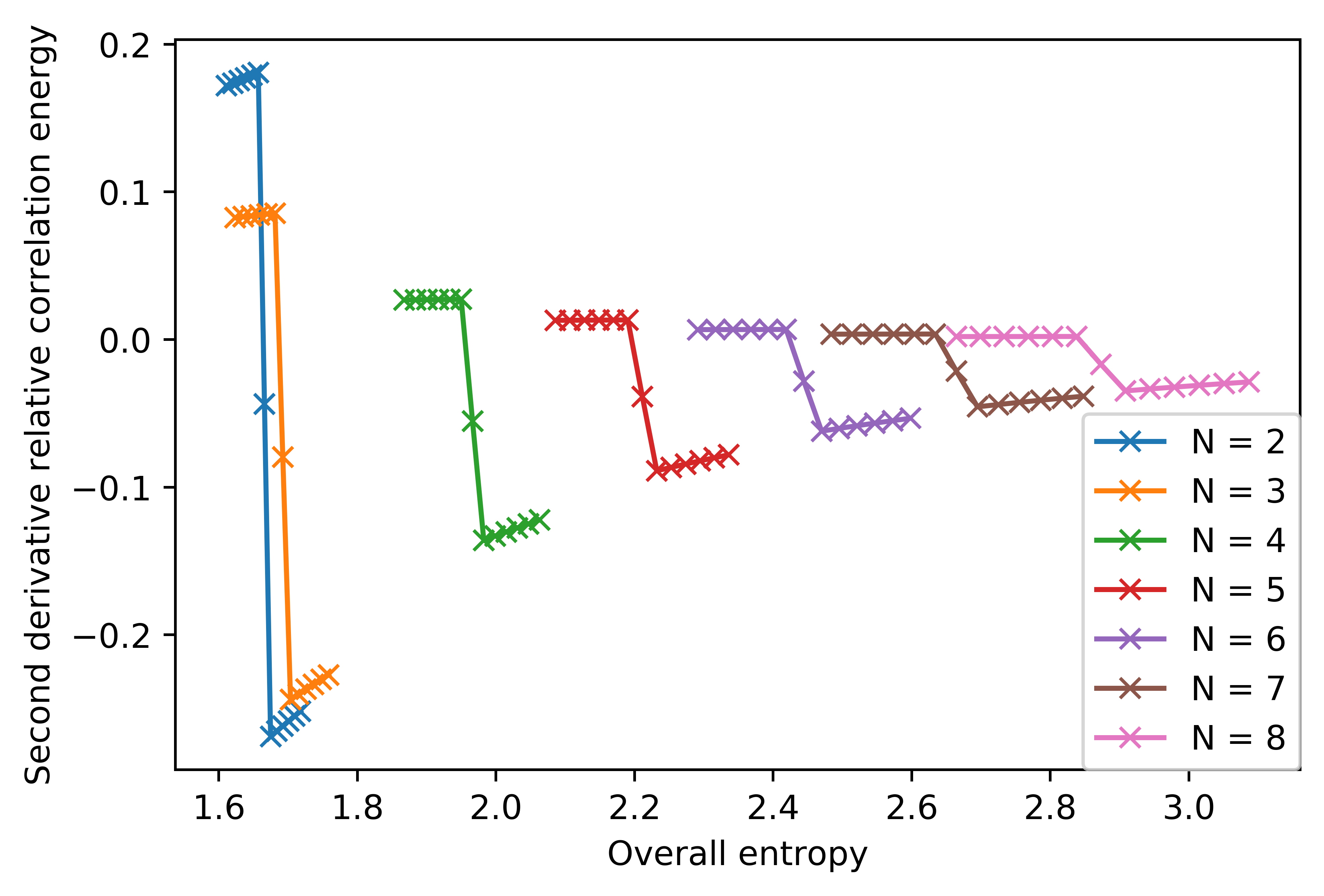}
\caption{Second derivative of the relative correlation energy (Fig. \ref{fig:2_lipkin_corr_energy_vs_overall_entropy}) for different values of $N$. We observe a discontinuity in $\chi = 1$, related to the phase transition of the model.}
\label{fig:second_derivative_2_lipkin_corr_energy_vs_overall_entropy}
\end{figure}

However, Fig. \ref{fig:second_derivative_2_lipkin_corr_energy_vs_overall_entropy} hides some subtleties. Although the phase transition at $\chi = 1$ is clear by the presence of the discontinuity, Fig. \ref{fig:second_derivative_2_lipkin_corr_energy_vs_overall_entropy} must not be interpreted as a `genuine phase transition indicator'. In a genuine phase transition we observe a change in the system's behaviour which becomes more evident  as the size of the system grows. In Fig. \ref{fig:second_derivative_2_lipkin_corr_energy_vs_overall_entropy} we see the opposite behaviour: the discontinuity is less abrupt when the system's size (the number of particles) is higher. This is due to the nature of the HF approximation: it is more accurate for higher values of $N$ \cite{ring_schuck}. In this way, we observe in Fig. \ref{fig:2_lipkin_corr_energy_vs_overall_entropy} lower values for the relative correlation energy as $N$ increases, and therefore the discontinuity in the second derivative is less abrupt.

We conclude from the previous discussion that the phase transition present in the HF solution is not only a `feature' of the mean field method but it also reflects a structural change in the exact wave function of the system. This statement is consistent with the fact  that the mean field approximation becomes more and more accurate as the number of particles in the system increase and also with the fact that the phase transition is better defined as the number of particles increases. To exemplify the later results, we show in Fig. \ref{fig:2_lipkin_onebody_vs_kappa} the averaged\footnote{The averaged overall entropy is simply the `overall entropy per particle', this is, $\frac{S_{ov}}{N}$.} overall entropy (for the exact solution) as a function of the interaction parameter $\chi$ for some values of $N$.

We observe a sudden change in the averaged one-body entropy when $\chi = 1$, which is sharper as the number of particles increases. As the value of the averaged one-body entropy is a measure of the correlations in the system we conclude that the spherical phase ($\chi<1$) corresponds to a low-correlated regime in the exact solution, while the deformed phase ($\chi\geq1$) corresponds to a high-correlated regime. Therefore, the behaviour of the overall entropy, which quantifies the total correlation, can help us to distinguish between different phases in the exact solution.

\begin{figure}
\includegraphics[width=0.5\textwidth]{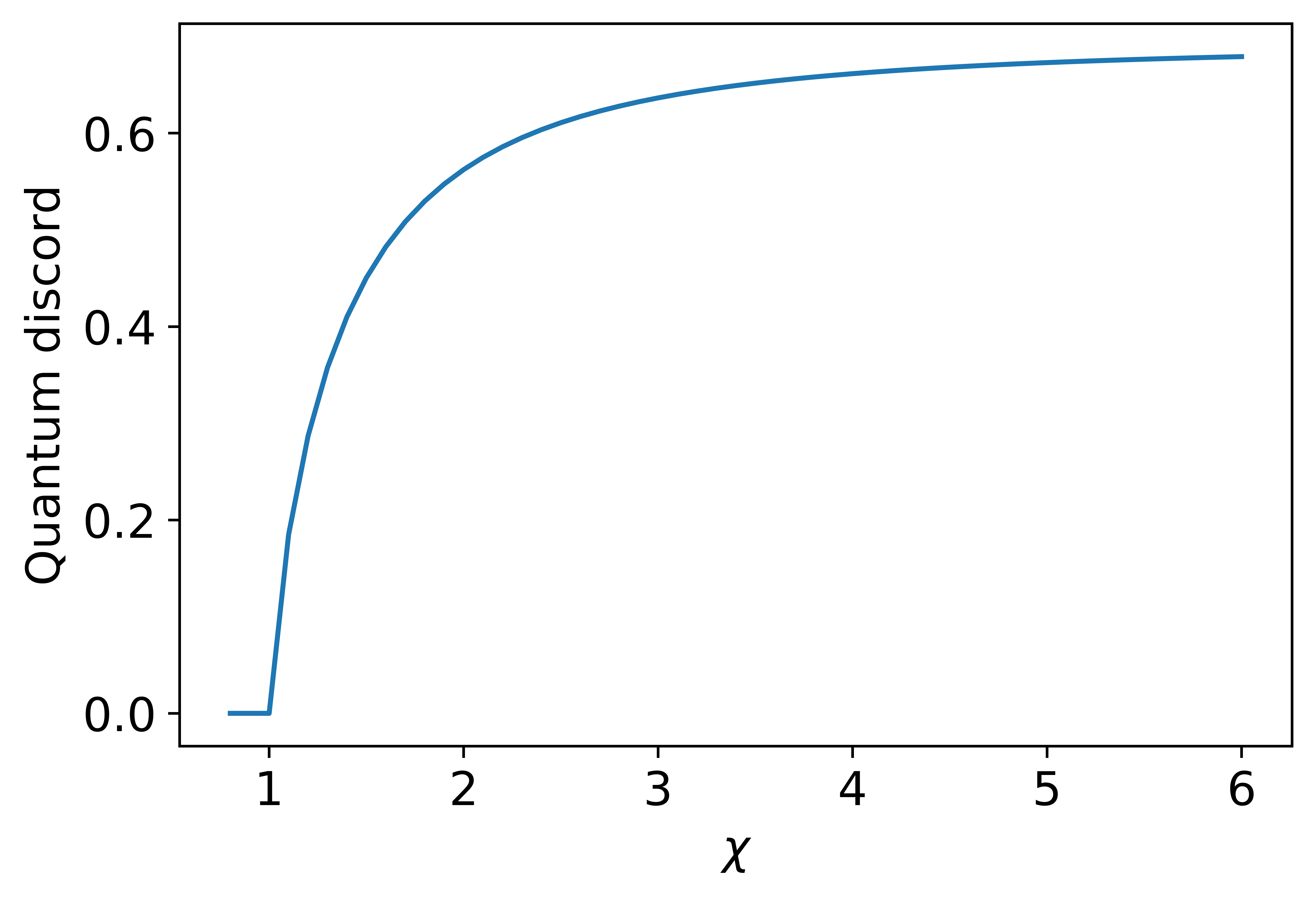}
\caption{Quantum discord between up and down levels of same quantum number $p$ (same degeneration level) for the HF ground state.}
\label{fig:2_lipkin_QD_HF_state}
\end{figure}

Other interesting quantity related to the overall entropy, which is computed from the mean field state, is the two-orbital quantum discord \cite{two_orbital_quantum_discord} between a couple of modes with same $p$ and opposite $\sigma$ for the HF ground state. Because of the symmetries of this model, the reduced density matrix of those modes are still pure (see Appendix \ref{app:purity}). For this reason, all the quantum correlations are entanglement and the quantum discord reduces to the entanglement entropy between modes. However, as we will see in Sec. \ref{sec:3_level_lipkin}, this will not be the case for the three level Lipkin model. If we plot the quantum discord  \cite{two_orbital_quantum_discord}
\begin{equation}
\label{eq:two_level_HF_quantum_discord}
    \delta(\sigma,p;-\sigma,p) = 
    \begin{cases}
        0, &\text{if } \chi\leq1  \\
        h(\chi), &\text{if } 1<\chi
    \end{cases}
\end{equation}
with $h(x) = -\frac{1}{2}(1-\frac{1}{x})\ln\frac{1}{2}(1-\frac{1}{x})-\frac{1}{2}(1+\frac{1}{x})\ln\frac{1}{2}(1+\frac{1}{x})$
as a function of the interaction parameter $\chi$ we obtain Fig. \ref{fig:2_lipkin_QD_HF_state}. 
 
As in Fig. \ref{fig:2_lipkin_onebody_vs_kappa}, we see clearly the quantum phase transition at $\chi = 1$. In fact, Fig. \ref{fig:2_lipkin_QD_HF_state} is very similar to Fig. \ref{fig:2_lipkin_onebody_vs_kappa} when the particle number is large. This is to be expected as the two orbital reduced state is pure and therefore the single orbital entropy represents the entanglement between the two orbitals. Thus, the overall entropy is twice the sum of the entanglement between the orbital pairs. On the other hand, if we use Eq. (\ref{eq:overall_entropy_and_onebody_entropy}) and we take into account that in the natural basis  $\langle a^\dagger_{-p}a_{-p}\rangle + \langle a^\dagger_{+p}a_{+p}\rangle = 1$, then $S_{ov}^{nat} = 2S(\gamma)$. For this reason, Fig. \ref{fig:2_lipkin_onebody_vs_kappa} and \ref{fig:2_lipkin_QD_HF_state} are almost the same in the limit $N\rightarrow{\infty}$. It is relevant to note that the quantum discord depicted in Fig. \ref{fig:2_lipkin_QD_HF_state} does not depend on the particle number since it is a `microscopic' quantity (i.e, it is defined between a couple of orbitals) of a mean-field state. However, we can see clearly the quantum phase transition in the behaviour of this quantity. Moreover, as discussed in \cite{entanglement_lipkin_model}, the nonzero quantum discord (entanglement in this model) showed in Fig. \ref{fig:2_lipkin_QD_HF_state} is a direct consequence of the symmetry breaking at the mean field level: for the exact ground state, the reduced density matrix for two levels with the same $p$ and opposite $\sigma$ does not have coherent elements and therefore entanglement. However, for the HF ground state, the reduced density matrix is pure and entangled.

\section{\label{sec:3_level_lipkin}Three level Lipkin model}

This model is a generalization \cite{Li1970} of the $N$-particle two level Lipkin model discussed in the previous section. There are  three energy levels in the model, each one with a $N$-fold degeneracy and, analogously to the two level Lipkin model, the interaction term can't change the degeneracy quantum number $p = 1,2,...,N$. If we assume that the interaction is the same for the three levels, which are equally spaced, we can write the Hamiltonian as
\begin{equation}
\label{eq:3_level_Lipkin_Hamiltonian}
    H = \epsilon(K_{22}-K_{00})-\frac{V}{2}(K^2_{10}+K^2_{20}+K^2_{21}+h.c)
\end{equation}
with
\begin{equation*}
    K_{\sigma\sigma '} = \sum_{p=1}^N c^\dagger_{\sigma p}c_{\sigma ' p}
\end{equation*}
As explained in \cite{Li1970,HOLZWARTH,Hagino2000} the exact ground state of Eq. (\ref{eq:3_level_Lipkin_Hamiltonian}) can be easily computed numerically in the basis $\ket{n_1,n_2}$, where $n_i$ is the number of particles in the $i$-th level. The basis elements are built upon the action of the operators $K_{10}^{n_1}$ and $K_{20}^{n_2}$ acting on the states with all the orbits in level 0 occupied. The given set of states is a basis to diagonalize $H$ because the operators $K_{\sigma\sigma '}$ are the generators of the algebra of $SU(3)$. 

\begin{figure}
\includegraphics[width=0.5\textwidth]{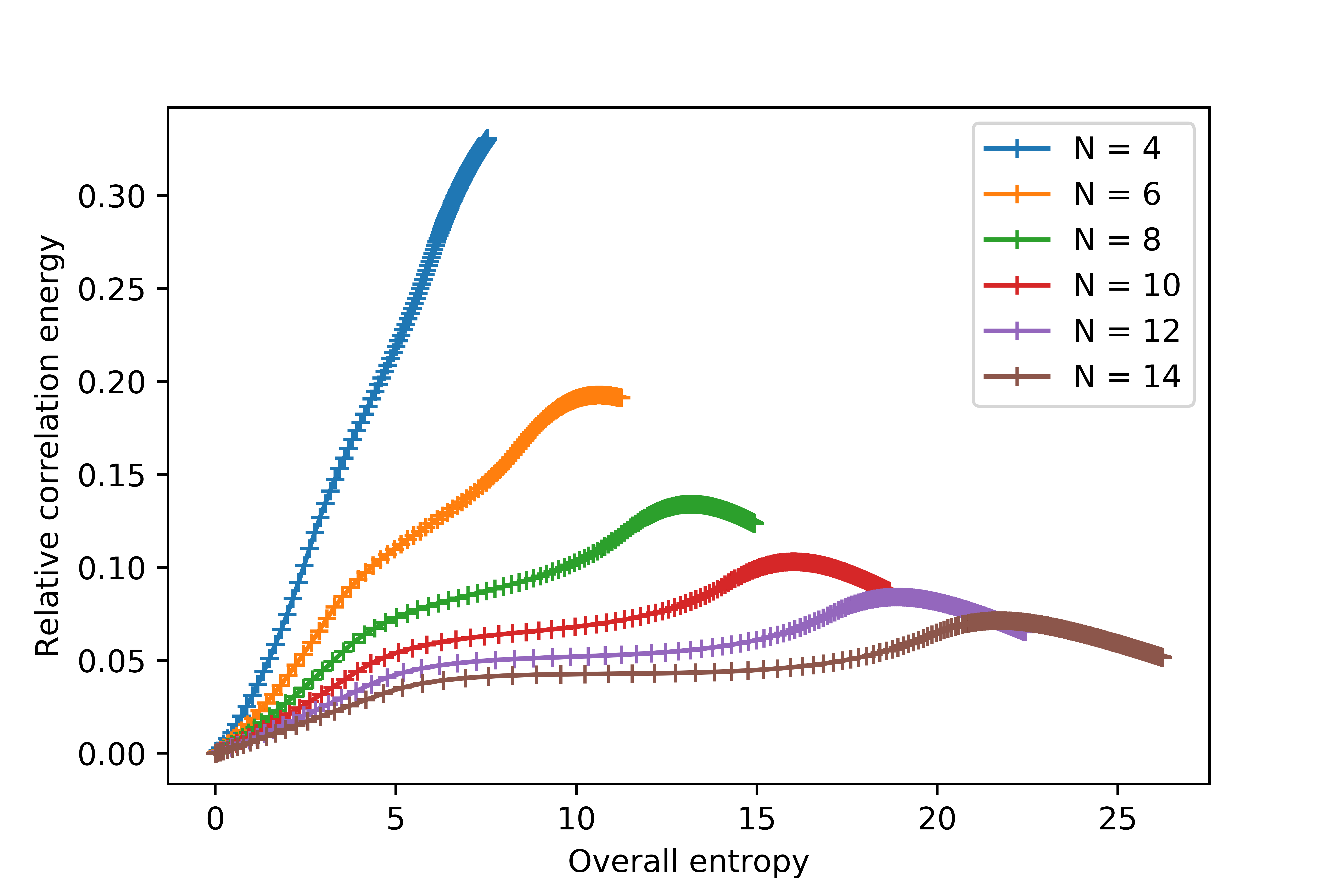}
\caption{Relative correlation energy as a function of the overall entropy, for the exact ground state of the three level Lipkin model.}
\label{fig:3_lipkin_corr_energy_vs_overall_entropy}
\end{figure}

\begin{figure}
\includegraphics[width=0.5\textwidth]{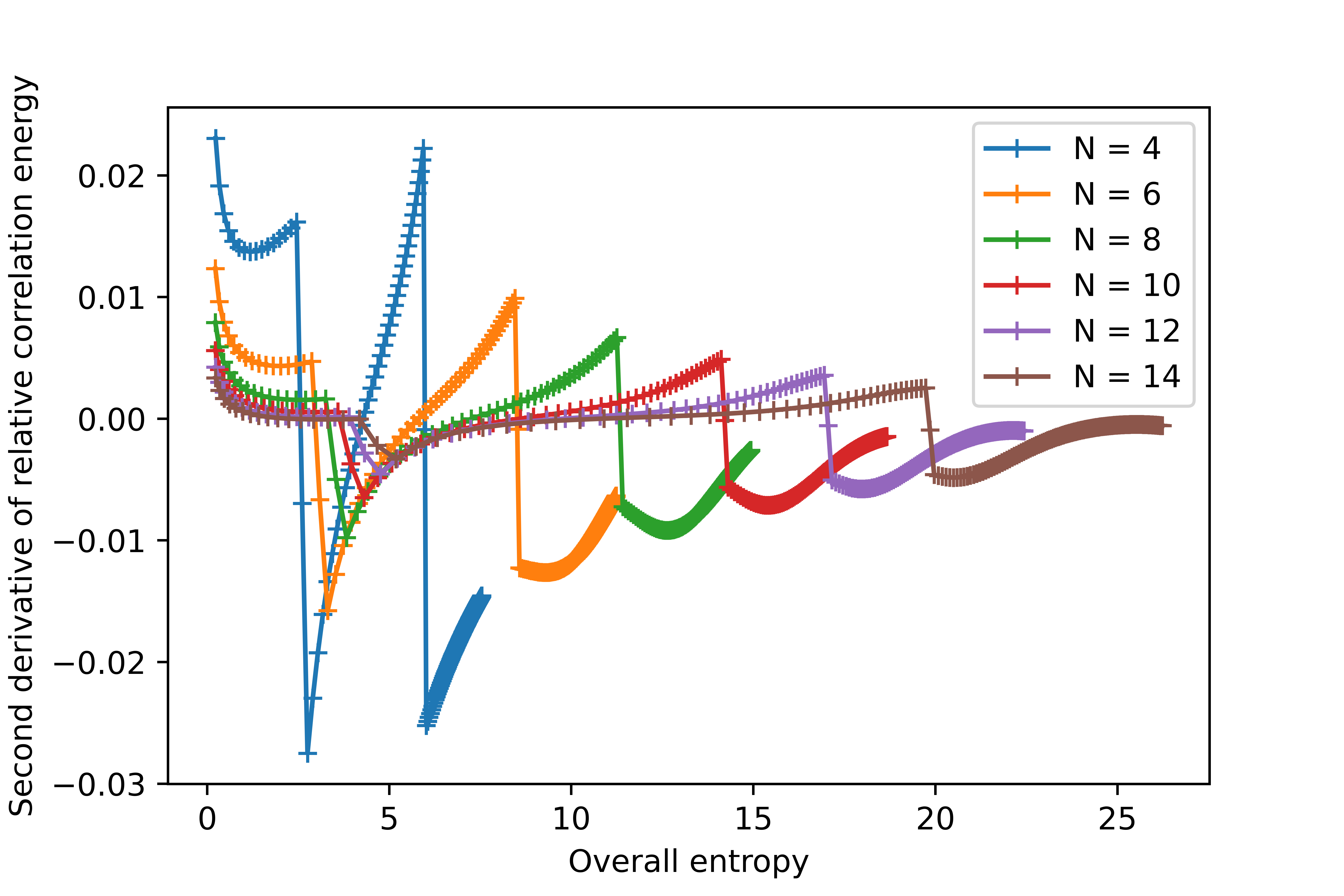}
\caption{Second derivative of Fig. \ref{fig:3_lipkin_corr_energy_vs_overall_entropy}. We observe a discontinuity in $\chi = 1$ and $\chi = 3$ related to the two phase transitions of this model.}
\label{fig:second_der_3_lipkin_corr_energy_vs_overall_entropy}
\end{figure}

If we compute the HF solution of the three level Lipkin model \cite{HOLZWARTH,Hagino2000}, it can be seen that there are two phase transitions, each one corresponding to the breaking of a level's symmetry. More precisely, the first phase transition is located at $\chi=1$ and corresponds to a parity-like breaking of the $\sigma = 1$ level, while the second one is located in $\chi = 3$ and corresponds to a parity-like breaking of the $\sigma = 2$ level. 
This behaviour is reflected in Fig. \ref{fig:3_lipkin_corr_energy_vs_overall_entropy} where the relative correlation energy as a function of the overall entropy, for the exact ground state is depicted in a similar way as in Fig \ref{fig:2_lipkin_corr_energy_vs_overall_entropy}.
When the system's correlation is low enough, the relative correlation energy grows quasi-linearly until reaching the second derivative discontinuity at $\chi = 1$ (Fig. \ref{fig:second_der_3_lipkin_corr_energy_vs_overall_entropy}). This is required in order to catch the maximum correlations as possible while maintaining the non-interacting ansatz. From this point on, the relative correlation energy increases more slowly until the the second quantum phase transition takes place at $\chi = 3$. From there on, the relative correlation energy decreases while the overall entropy increase reflecting the fact the mean field solution approximates better the exact solution. Finally, as in Fig. \ref{fig:2_lipkin_corr_energy_vs_overall_entropy}, the overall entropy saturates.

As discussed in Sec. \ref{sec:2_level_lipkin}, the relative correlation energy acquire lower values as the particle number increases, in agreement with the general idea that the mean field picture increases its accuracy in the thermodinamic limit (infinite number of particles). This is the reason why the discontinuity in the second derivative depicted in Fig. \ref{fig:second_der_3_lipkin_corr_energy_vs_overall_entropy} for different values of particle number $N$ is less and less pronounced as  $N$ gets higher and higher. The behaviour is essentially the same as in the two level Lipkin model except for the double quantum phase transition.

\begin{figure}
\includegraphics[width=0.5\textwidth]{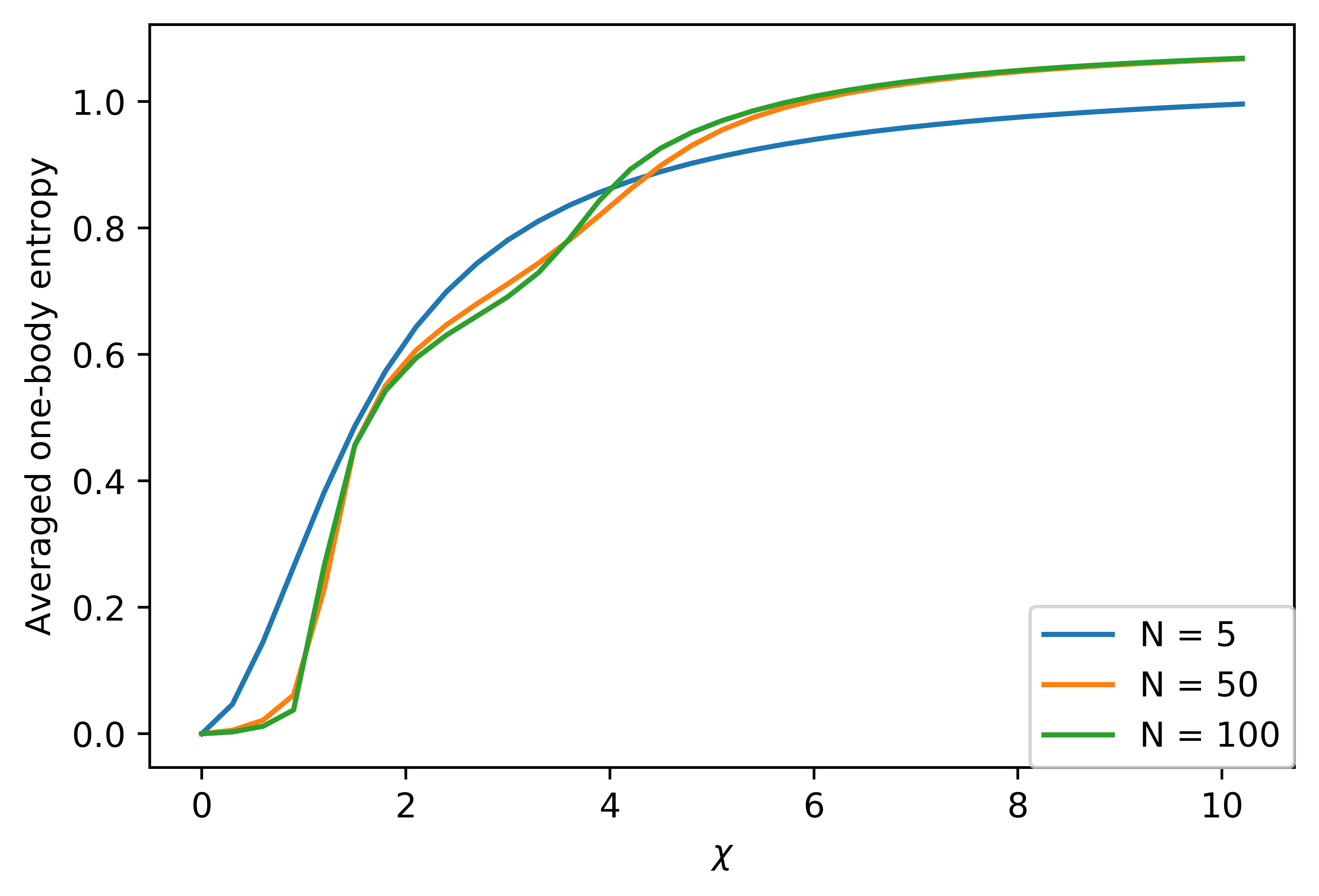}
\caption{Averaged one body entropy of the exact ground state as a function of the parameter $\chi$.}
\label{fig:3_lipkin_onebody_vs_kappa}
\end{figure}

\begin{figure}
\includegraphics[width=0.5\textwidth]{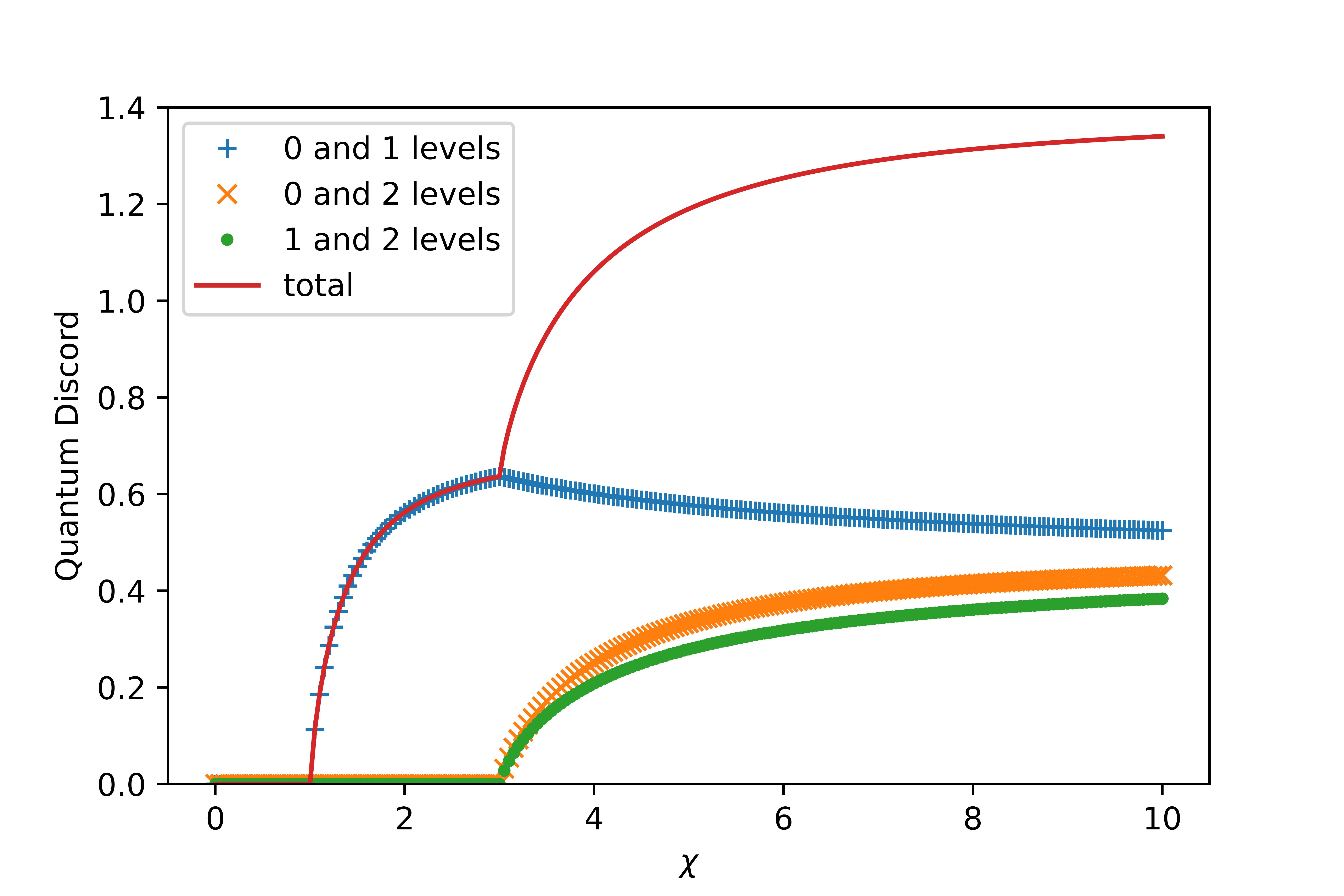}
\caption{Quantum discord between levels of the same degeneration number $p$ for the HF ground state. The sum of the quantum discord between the three different configurations is represented with the solid red line.}
\label{fig:3_lipkin_QD_HF_state}
\end{figure}

Following the same analysis as in Sec. \ref{sec:2_level_lipkin}, we can study the quantum phase transitions through the overall entropy as a function of the interaction parameter (Fig. \ref{fig:3_lipkin_onebody_vs_kappa}), and compare it with the quantum discord between levels with different $\sigma$ and same $p$ (Fig. \ref{fig:3_lipkin_QD_HF_state}).

The behaviour of the overall entropy is very similar in both models. If $N$ is small (for example $N=5$ in Fig. \ref{fig:3_lipkin_onebody_vs_kappa}) the shape of the overall entropy is almost the same for the two and three Lipkin models: since a quantum phase transition is a global property there is no difference between phases. However, as the particle number increases, the distinction between the three regions (spherical phase in $\chi\leq1$, first parity break in $1\leq\chi\leq3$ and second parity break in $3\leq\chi$) is sharper, and the differences between the two and three level Lipkin models arises. They can be clearly observed through the quantum discord between levels of the same degeneration number $p$ for the HF ground state (Fig. \ref{fig:3_lipkin_QD_HF_state})\footnote{Unlike in the two level Lipkin model, here the reduced state is in general mixed, and we can't compute the entanglement as in Eq. (\ref{eq:vn_entropy}).}. 
For the spherical region, there is no quantum correlation between any level since the HF orbitals are related to the original ones through the identity matrix. That is, the mean field state is simply the non interacting ground state of Eq. (\ref{eq:3_level_Lipkin_Hamiltonian}). When the first symmetry breaking occurs, the quantum correlations between states with $\sigma = 0$ and $1$ increases abruptly, while it remains zero between $\sigma = 0$ and $2$ and $\sigma = 1$ and 2. Indeed, its value is exactly the same as the two level Lipkin model (see Eq. (\ref{eq:two_level_HF_quantum_discord}) and (\ref{eq:3_level_Lipkin_QD_01})). Since the $\sigma = 2$ level remains unfilled (the $\sigma = 0$ and $1$ levels are mixed while the $\sigma = 2$ level is not), there is no difference between quantum correlations of the two and three level Lipkin models within the mean field description. However, when the second symmetry breaking occurs, the three $\sigma$ levels are completely mixed. The quantum correlations between $\sigma = 0$ and $1$ levels spontaneously decreases due to the redistribution of the occupation between all levels, while the quantum discord between $\sigma = 0$ and $2$ and $\sigma = 1$ and $2$ grows in a very similar fashion (being the quantum correlations between $\sigma = 1$ and $2$ always lower). Finally, it is interesting to note that if we compare the sum of the quantum discord between the three possible orbital combinations (solid red line in Fig. \ref{fig:3_lipkin_QD_HF_state}) with the one-body entropy in Fig. \ref{fig:3_lipkin_onebody_vs_kappa}, we see that if $N$ is high enough, both line's shapes follow the same `double jump' trend.

\section{\label{sec:conclusions}Conclusions}

The relative correlation energy has been typically used in order to quantify the amount of correlation in a state, since it is defined as the relative difference between the exact energy and the mean-field one. On the other hand, with the fast growth in the last decades of the quantum information field, there are a variety of methods nowadays in order to quantify the correlation in a system in terms of their subsystems. An example are the entanglement entropy, mutual information, quantum discord or the one-body entropy. In this work we have analyzed the relative correlation energy and some quantum information measures in the context of the two and three level Lipkin model. We found that the relative correlation energy is not a good estimator of the total correlation of a system, but it is a good estimator of the accuracy of the mean field approximation. Comparing the overall entropy (which is a measure of the total system's correlation under the quantum information context) and the relative correlation energy we don't find quasi-linear or monotonously increasing behaviour. Indeed, we find regions in the parameter space in which the overall entropy grows but the relative correlation energy tends to decrease, and regions in which both tend to grow. Those regions are defined by quantum phase transitions, which can be analyzed by computing the quantum discord between orbitals at HF level, without the need of computing the exact ground state.

Future work includes the analysis of different models, both analytically or numerically solvable, such as $N$-level Lipkin, picket fence or single-$j$ shell models. Also, a more exhaustive analysis can be performed in more complex systems by computing the quantum discord between bigger orbital subsystems of interest, or extending the mean field picture to a quasiparticle vacuum.

\acknowledgments{}
The authors want to thank the Madrid regional government, Comunidad 
Aut\'onoma de Madrid, for the project Quantum Information Technologies: 
QUITEMAD-CM P2018/TCS-4342.
The  work of LMR was supported by Spanish Ministry 
of Economy and Competitiveness (MINECO) Grants No. 
PGC2018-094583-B-I00. We would like to thank Jorge Tabanera for enlightening discussions.

\appendix
\section{\label{app:purity} Purity of the two orbital reduced density matrix for the Hartree-Fock ground state of the two level Lipkin model}

In this section we will compute the purity of the two orbital reduced density matrix for the HF ground state of the two level Lipkin model. As explained in \cite{agassi}\footnote{Here the authors work with the Agassi model, which is an extension of the two level Lipkin model.} we can write the one-body density matrix of the HF ground state as
\begin{equation*}
\gamma_{\sigma p,\sigma ' p'} = 
\begin{cases}
    \frac{1}{2}(1-\sigma \cos\varphi)\delta_{p,p'}, &\text{if } \sigma = \sigma '  \\
    -\frac{1}{2}\sin\varphi\delta_{p,p'}, &\text{if } \sigma = -\sigma '
\end{cases}
\end{equation*}
with 
\begin{equation*}
\cos\varphi = 
\begin{cases}
    1, &\text{if } \chi\leq 1  \\
    \frac{1}{\chi}, &\text{if } \chi>1
\end{cases}
\end{equation*}
Following the results in \cite{two_orbital_quantum_discord}, the two orbital reduced density matrix is

\begin{equation*}
    \rho^{(A,B)} = \frac{1}{2}
    \begin{pmatrix}
        0 & 0 & 0 & 0 \\
        0 & 1+\cos\varphi & -\sin\varphi & 0 \\
        0 & -\sin\varphi & 1-\cos\varphi & 0 \\
        0 & 0 & 0 & 0 \\
    \end{pmatrix}
\end{equation*}
whose eigenvalues are $0$ and $1$.

\section{\label{app:3_level_Lipkin_QD} Quantum discord for the Hartree-Fock state in the three level Lipkin model}

In this section we will briefly develop the analytic expression for the two orbital quantum discord in the HF state of the three level Lipkin model. Following reference \cite{two_orbital_quantum_discord}, we only need to compute the one-body elements and the two-body diagonal elements for each orbital. If we assume that the system is in the HF ground state, i.e, $\ket{HF} = \prod_{q = 1}^N a^\dagger_{0q}\ket{0}$ (with $\ket{0}$ the vacuum state) then, using Wick's theorem,
\begin{equation*}
\begin{split}
    \bra{HF}c^\dagger_{\alpha i}c_{\beta j}\ket{HF} &= U^\dagger_{\alpha 0}U_{0\beta}\delta_{ij} \\
    \bra{HF}c^\dagger_{\alpha i}c^\dagger_{\beta j}c_{\beta j}c_{\alpha i}\ket{HF} &= \abs{U_{0\alpha}U_{0\beta}}^2(1-\delta_{ij})
\end{split}
\end{equation*}
with $a^\dagger_{\alpha i} = \sum_{\beta = 0}^2 U_{\alpha\beta}c^\dagger_{\beta i}$ and $U U^\dagger = 1$. Following the results in \cite{HOLZWARTH,Hagino2000}, the mean field solution can be written as
\begin{equation}
\label{eq:3_Lipkin_change_of_basis}
    U = 
    \begin{pmatrix}
        c_\alpha & c_\beta s_\alpha & s_\beta s_\alpha \\
        -c_\beta s_\alpha & 1+c^2_\beta(c_\alpha-1) & s_\beta c_\beta(c_\alpha-1) \\
        -s_\beta s_\alpha & s_\beta c_\beta(c_\alpha-1) & 1+s^2_\beta(c_\alpha-1) 
    \end{pmatrix}
\end{equation}
with $c_\alpha \coloneqq \cos\alpha$, $s_\alpha \coloneqq \sin\alpha$ and
\begin{equation*}
\cos^2\alpha = 
    \begin{cases}
        1, &\text{if } \chi\leq1  \\
        \frac{1}{2}(1+\frac{1}{\chi}), &\text{if } 1<\chi\leq3  \\
        \frac{\chi+3}{3\chi}, &\text{if } 3<\chi  \\
    \end{cases}
\end{equation*}
\begin{equation*}
\cos^2\beta = 
    \begin{cases}
        1, &\text{if } 1<\chi\leq3  \\
        \frac{1}{2}(\frac{3}{2\chi-3}+1), &\text{if } 3<\chi  \\
    \end{cases}
\end{equation*}
Using those results and Eq. (5) in \cite{two_orbital_quantum_discord}, we easily obtain the analytic expression for the quantum discord between any orbital pair:
\begin{widetext}
\begin{equation}
\label{eq:3_level_Lipkin_QD_01}
\begin{split}
\delta(0,p;1,p) =&
    \begin{cases}
        0, & \text{if } \chi\leq1  \\
        s(\frac{1}{2}(1+\frac{1}{\chi})) + s(\frac{1}{2}(1-\frac{1}{\chi})), &\text{if } 1<\chi\leq3  \\
        -s(\frac{2}{3}+\frac{1}{\chi}) + s(\frac{1}{3}+\frac{1}{\chi})+s(\frac{1}{3}), &\text{if } 3<\chi  \\
    \end{cases} \\
\delta(0,p;2,p) =&
    \begin{cases}
        0, & \text{if } \chi\leq3  \\
        s(\frac{1}{3}+\frac{1}{\chi}) + s(\frac{1}{3}-\frac{1}{\chi})-s(\frac{2}{3}), & \text{if } 3<\chi  \\
    \end{cases} \\
\delta(1,p;2,p) =&
    \begin{cases}
        0, &\text{if } \chi\leq3  \\
        -s(\frac{2}{3}-\frac{1}{\chi}) + s(\frac{1}{3}-\frac{1}{\chi})+s(\frac{1}{3}), &\text{if } 3<\chi  \\
    \end{cases}
    \end{split}
\end{equation}
\end{widetext}
with $s(x) = -x\log x$.

\bibliography{bibliography}

\end{document}